\documentclass{article}
\usepackage[most]{tcolorbox}
\usepackage{subcaption}
\usepackage{iclr2027_conference,times}

\usepackage{amsmath,amsfonts,bm}

\def\eqref#1{equation~\ref{#1}}

\def\1{\bm{1}}

\DeclareMathAlphabet{\mathsfit}{\encodingdefault}{\sfdefault}{m}{sl}
\SetMathAlphabet{\mathsfit}{bold}{\encodingdefault}{\sfdefault}{bx}{n}

\usepackage{pifont}
\usepackage{amsmath,amssymb}
\usepackage{booktabs}
\usepackage{graphicx}
\usepackage{multirow}
\usepackage[most]{tcolorbox}
\usepackage{hyperref}
\usepackage{url}
\usepackage{dsfont}

\usepackage{etoolbox}
\makeatletter
\patchcmd{\iclrfinalcopy}{}{}{}{}
\makeatother

\title{SymbolicLM: Training Language Models as Symbolic Regressors}

\author{
Jun Yao$^{1,2}$,
Yingfan Hua$^{3}$,
Ruikun Li$^{2}$,
Shixiang Tang$^{2}$,
Bin Liu$^{1\dagger}$,
Wanli Ouyang$^{2,3}$,
Yan Lu$^{2,3\dagger}$
\\
$^{1}$University of Science and Technology of China\\
$^{2}$Shanghai Artificial Intelligence Laboratory\\
$^{3}$ The Chinese University of Hong Kong\\
\texttt{\{flowice@ustc.edu.cn, luyan@pjlab.org.cn\}}
}

\begin{document}

\maketitle

\maketitle
\begingroup\def\thefootnote{$\dagger$}\footnotetext{Corresponding authors.}\endgroup

\begin{abstract}
Large Language Models (LLMs) have demonstrated strong capabilities in scientific reasoning, drawing growing attention to their potential in scientific discovery. However, scientific discovery fundamentally originates from deriving formal laws directly from observational data, known as Symbolic Regression (SR). This task poses a critical challenge to LLMs, stemming from an inherent tension between pre-trained LLMs’ proficiency in approximate reasoning, rooted in probabilistic text generation, and the high-precision demands of SR tasks. While recent methods employ complex external scaffolds to mitigate this limitation, such an iterative agentic paradigm remains computationally inefficient and fundamentally separates the model’s internal scientific knowledge from the symbolic regression process. To address this problem, we propose to directly equip LLMs with precise symbolic understanding and the ability to induce laws from raw observational data. In this paper, we introduce \textbf{PhysSymbArena}, a large-scale benchmark, comprising more than $160{,}000$ diverse equations and $1.8$B tokens of numerical symbolic data and physical description, to support post-training and extensive evaluation. Building on PhysSymbArena, we propose \textbf{SymbolicLM}, in which the symbolic regression capability of LLMs is systematically enhanced during post-training through \textbf{mathematical and physical supervision}. At inference time, we introduce a SymbolicSGA Refinement framework that uses quantitative fitting feedback to iteratively refine and recombine promising symbolic structures, enabling the model to correct structural errors and progressively recover more accurate governing equations. Experiments on multiple symbolic regression benchmarks demonstrate that SymbolicLM substantially outperforms representative symbolic regression methods in structural recovery while maintaining competitive numerical fitting performance. These results demonstrate that symbolic regression can be explicitly learned and strengthened as an intrinsic capability of LLMs.

\end{abstract}

\section{Introduction}
\label{sec:introduction}
Large language models (LLMs) have demonstrated strong capabilities across a wide range of scientific tasks, including scientific question answering, literature understanding, mathematical reasoning, and scientific text generation ~\citep{taylor2022galactica,lewkowycz2022solving,wang2023scibench}. Yet, genuine scientific discovery extends beyond reasoning over established knowledge and predefined problems; it fundamentally requires deriving exact, formal laws directly from observational data. A canonical formulation of this problem is symbolic regression (SR), which aims to discover governing equations that describe the data. Despite their reasoning power, general-purpose LLMs face distinct challenges in solving symbolic regression.

The limitations of LLMs in symbolic regression fundamentally stem from an inherent tension between their proficiency in approximate reasoning and the high-precision demands of the task. Specifically, these high-precision demands arise from the strict objective of the problem: given a set of numerical observations $\mathcal{D}=\{(\mathbf{x}_i,y_i)\}_{i=1}^{N}$, the model must identify an analytical expression $f$ that consistently satisfies $y_i \approx f(\mathbf{x}_i)$ across all data points. While pre-trained LLMs excel at probabilistic next-token generation, where approximate plausibility is sufficient, evaluating a mathematical law admits no margin for structural error. Even a minor mistake in an operator, exponent, or variable dependency is amplified across the input domain, causing $f(\mathbf{x}_i)$ to diverge substantially from the observed $y_i$. To mitigate this limitation, existing methods predominantly turn to external scaffolds, yet this reliance incurs prohibitive computational costs while still yielding suboptimal accuracy.

\begin{figure*}[t]
    \centering
\includegraphics[width=0.95\textwidth]{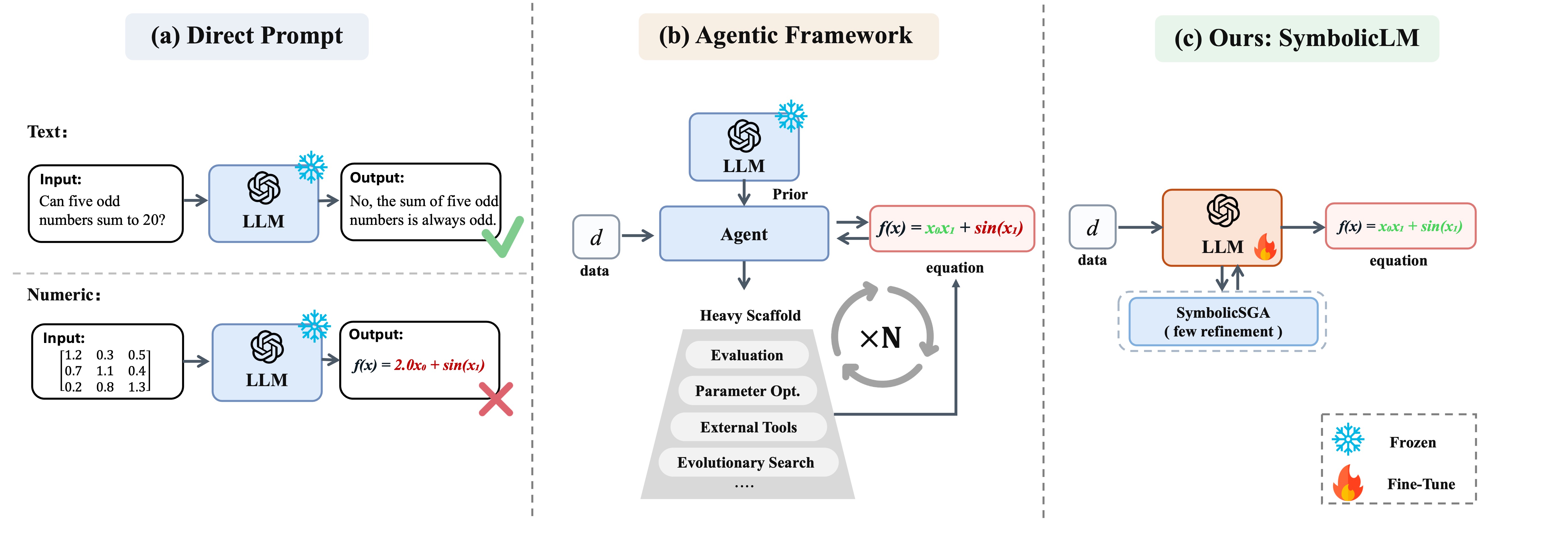}
    \caption{Comparison of different frameworks for symbolic equation discovery with LLMs.}
    \label{fig:figure1}
\end{figure*}

Existing scaffold-based methods typically rely on complex agentic workflows, where the LLM proposes candidate formulas and external modules evaluate numerical fitting~\citep{ma2024llm,grayeli2024symbolic,shojaee2025llm}. However, this division of labor introduces a fundamental feedback bottleneck. In-context prompts cannot effectively translate scalar numerical residuals into precise structural algebraic edits; the model struggles to discern whether a poor fit stems from incorrect functional forms or unoptimized parameters. Consequently, the search easily falls into local optima or constructs ungrounded expressions, failing to recover the true mathematical skeleton. This blind, multi-round trial and error can take up to hours for a single equation, yet still yields structurally flawed results. To overcome this limitation, in this paper, we propose to directly instill symbolic regression capabilities into LLMs, enabling them to discover governing laws from observational data as an intrinsic capability.As illustrated in Figure~\ref{fig:figure1}.

To instill intrinsic symbolic regression capabilities into LLMs, we introduce a unified framework spanning physically grounded data synthesis, mathematical-physical tuning, and lightweight inference. First, we introduce \textbf{PhysSymbArena}, a large-scale benchmark featuring physically grounded synthetic data. By systematically coupling over $160{,}000$ equations ($1.8$B tokens) with physical quantities, dimensional constraints, and variable descriptions, it enables LLMs to exploit internal scientific priors rather than treating observations as ungrounded numbers. Second, building on this foundation, we develop \textbf{SymbolicLM} through a dedicated Mathematical-Physical Tuning (MPT), realized via a reinforcement fine-tuning framework. Going beyond token-level cross-entropy, this stage directly aligns policy through tailored rewards covering symbolic skeletons, physical constraints, and syntactic validity. This allows the model to produce valid, algebraically equivalent expressions without being confined to rigid token matching. Finally, at inference time, we design the SymbolicSGA refinement framework, a lightweight refinement loop. Empowered by the model's intrinsic structural awareness, SymbolicSGA eliminates the need for blind exploration, requiring only a few iterations of quantitative feedback to rectify residual errors and pinpoint exact governing equations.


Experiments demonstrate that explicitly training and aligning LLMs for symbolic regression substantially improves their ability to infer analytical structures from numerical observations. On PhysSymbArena, SymbolicLM significantly outperforms general-purpose LLM baselines. Moreover, on the external SRBench-Feynman benchmark, our method achieves strong performance against representative symbolic regression approaches, reaching an exact symbolic skeleton accuracy of 57.98\% while maintaining competitive numerical fitting accuracy. These results further demonstrate that the symbolic regression capability learned from PhysSymbArena can generalize beyond the training distribution to established symbolic regression benchmarks.

Our main contributions are summarized as follows:

\begin{itemize}
    \item We introduce \textbf{PhysSymbArena}, a large-scale numerical--symbolic platform containing more than $160{,}000$ diverse equations and $1.8$ billion tokens, designed to support both the training and systematic evaluation of LLM symbolic regression capabilities.

    \item We propose \textbf{SymbolicLM}, which combines supervised fine-tuning with GRPO-based symbolic and physical feedback to directly align LLM generation with symbolic regression objectives, including syntactic validity, physical consistency, analytical structure, symbolic equivalence, and numerical accuracy.

    \item We introduce an SymbolicSGA refinement mechanism that enables the model to continuously revise its symbolic hypotheses according to quantitative experimental feedback. Extensive experiments on three symbolic regression benchmarks show that SymbolicSGA achieves strong performance in symbolic structure recovery and numerical fitting accuracy.
\end{itemize}

\section{Related Work}
\label{sec:related_work}
\paragraph{Symbolic regression.}
Symbolic regression has traditionally been formulated as a search problem over the discrete space of mathematical expressions. Genetic programming and evolutionary approaches construct and modify expression trees through operations such as mutation, crossover, and selection, while reinforcement-learning methods such as Deep Symbolic Regression (DSR) learn policies for generating high-reward expressions~\citep{petersen2019deep,cranmer2023interpretable}. Although effective, these approaches typically perform substantial task-specific search for each new problem. A complementary line of work seeks to amortize this search process by learning numerical-to-symbolic mappings from large collections of synthetically generated equations. NeSymReS trains a Transformer to infer symbolic expressions from numerical observations and combines neural prediction with subsequent symbolic search~\citep{biggio2021neural}, while E2E Symbolic Regression directly decodes complete analytical expressions from observed data~\citep{kamienny2022end}. SNIP further improves numerical--symbolic representation learning through large-scale pretraining~\citep{meidani2024snip}.These methods demonstrate that numerical-to-symbolic mappings can be effectively learned from large-scale synthetic data. However, they are designed specifically for symbolic regression, whereas whether general-purpose LLMs can be explicitly trained to perform symbolic regression directly from numerical observations remains less explored.

\paragraph{LLMs for symbolic regression.}
The emergence of general-purpose LLMs has introduced a new paradigm for symbolic regression by leveraging the mathematical and scientific priors acquired during large-scale pretraining. Rather than training models specifically for numerical-to-symbolic prediction, recent approaches typically use LLMs to propose candidate equations or scientific hypotheses and combine them with external search, numerical optimization, or iterative refinement. ICSR adopts an iterative prompting strategy, guiding the LLM to propose and revise functional forms according to numerical fitting errors~\citep{merler2024context}. LLM-SR represents candidate equations as programs and combines LLM-generated hypotheses with evolutionary search and numerical parameter optimization~\citep{shojaee2025llm}. SGA similarly employs LLMs to reason over discrete scientific hypotheses while delegating continuous parameter optimization to differentiable procedures~\citep{ma2024llm}, whereas LaSR further improves hypothesis search by extracting reusable concepts from previously discovered high-performing candidates~\citep{grayeli2024symbolic}. These methods demonstrate that the scientific priors encoded in LLMs can substantially improve equation discovery. However, under this paradigm, LLMs primarily provide candidate structures or search priors, while a substantial portion of the symbolic regression process still relies on external search and optimization mechanisms. Consequently, the LLM itself is not explicitly trained to directly perform symbolic regression from numerical observations.

\section{PhysSymbArena: Dataset Construction}
\begin{figure*}[t]
    \centering
\includegraphics[width=0.95\textwidth]{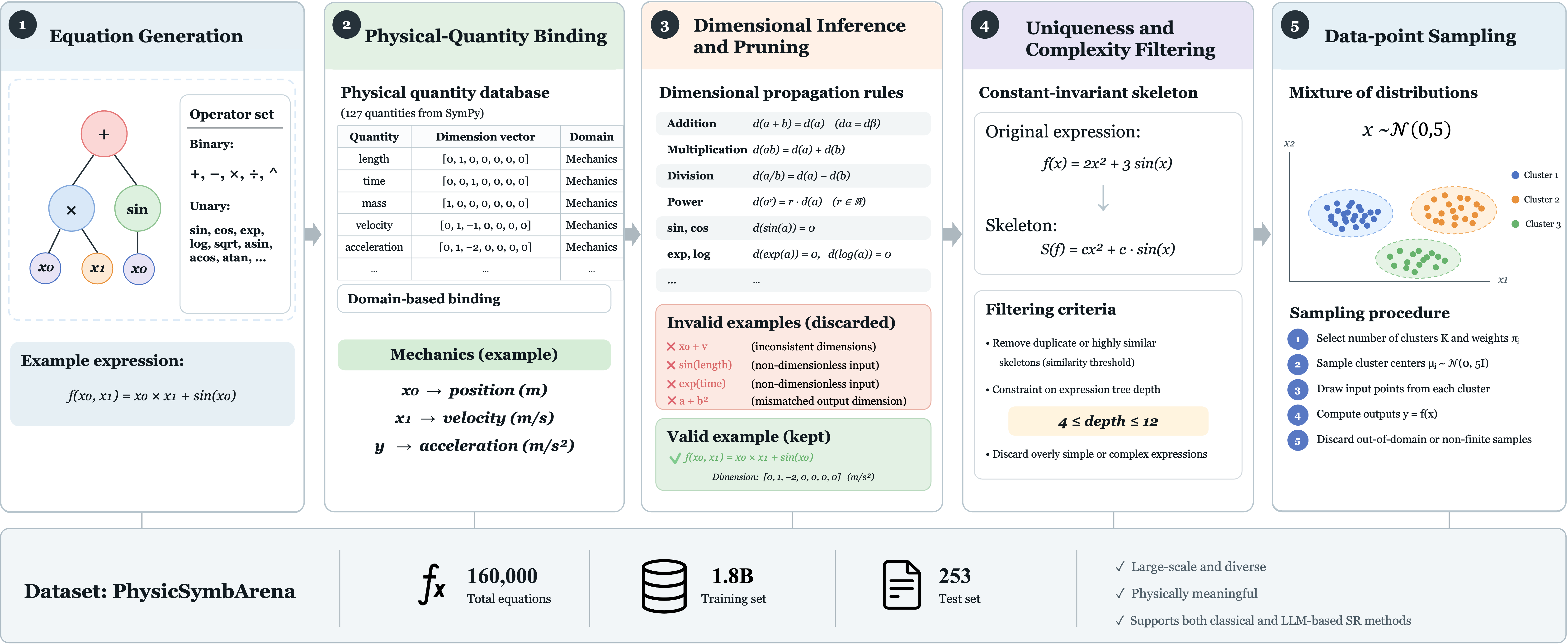}
    \caption{Overview of our proposed method.}
    \label{fig:overview}
\end{figure*}

We introduce \textbf{PhysSymbArena}, a large-scale symbolic regression dataset comprising $160{,}000$ analytical expressions with physical descriptions and numerical observations, totaling $1.8$B tokens. Each sample is formalized as a tuple $(\mathcal{D}, \mathcal{C}, f)$, where $f:\mathbb{R}^{D}\rightarrow\mathbb{R}$ denotes the target expression ($y=f(\mathbf{x})$), $\mathcal{C}$ encapsulates physical semantics and dimensional metadata, and $\mathcal{D}=\{(\mathbf{x}_k,y_k)\}_{k=1}^{N}$ represents empirical observations. As illustrated in Figure~\ref{fig:overview}, dataset construction proceeds across two steps: (1)~\textit{numerical equation generation and observation sampling}, and (2)~\textit{physical grounding and dimensional validation}. Comprehensive details could be seen in Appendix.


\paragraph{Numerical Equation Generation and Observation Sampling.}
We synthesize mathematical expressions and their corresponding numerical observations through a randomized tree-based simulation pipeline~\citep{kamienny2022end,meidani2024snip}. Specifically, candidate expressions $f$are constructed as trees whose depths and mathematical operators are randomly drawn from predefined libraries to introduce diverse structural complexities. For each generated expression, we sample input vectors across the domain and evaluate$f$to obtain the corresponding responses, formulating the numerical observation set$\mathcal{D}$.

\paragraph{Physical Grounding and Dimensional Validation.}
To endow abstract expressions with real-world physical semantics, we curate a dictionary of $127$ physical quantities across domains such as mechanics, electromagnetism, and thermodynamics from SymPy. For each expression, candidate physical quantities from a shared domain are randomly assigned to the input variables $\mathbf{x}$ and the target output $y$. We then perform bottom-up dimensional propagation along the expression tree under standard SI rules: addition and subtraction require identical dimensions, transcendental functions require dimensionless arguments, and the propagated root dimension must match that of $y$. Assignments that violate dimensional consistency are discarded and resampled, ensuring that the resulting physical metadata $\mathcal{C}$ strictly complies with physical laws.

\begin{table*}[ht!]
\caption{Comparison of symbolic regression benchmarks. SymbArena distinguishes itself through its massive scale, the inclusion of a train/test split, and its support for both traditional and LLM-based methods.}
\centering
\resizebox{0.8\textwidth}{!}{%
\renewcommand{\arraystretch}{1}
\begin{tabular}{@{}lcccc@{}}
\toprule
\textbf{Benchmarks} 
& \textbf{Numbers of Equations} 
& \textbf{Train \& Test} 
& \textbf{Supported Methods} \\
\midrule

Nguyen 
& 12 
& \ding{56} 
& Traditional only \\

Rational 
& 3 
& \ding{56} 
& Traditional only \\

SRbench 
& 252 
& \ding{56} 
& Traditional only \\

LLM-SRbench 
& 239 
& \ding{56} 
& LLM-based only \\

\midrule

SymbArena 
& 160,000 
& \ding{52} 
& Both \\

\bottomrule
\end{tabular}
}
\label{t1}
\end{table*}
\paragraph{Dataset Statistics}

Table~\ref{t1} compares PhysSymbArena with several widely adopted symbolic regression benchmarks, including Nguyen~\cite{uy2011semantically}, Rational~\cite{krawiec2013approximating}, SRBench~\cite{la2021contemporary}, and LLM-SRBench~\cite{shojaee2025llm}. PhysSymbArena contains $160{,}000$ equations, substantially exceeding the scale of existing benchmark datasets. Unlike benchmarks designed solely for evaluation, PhysSymbArena supports both model training and standardized testing. Moreover, PhysSymbArena supports both conventional symbolic regression algorithms and LLM-based approaches, providing a unified platform for training and evaluating different SR paradigms.




\section{SymbolicLM}
\label{sec:method}

\subsection{Overview}

Building on PhysSymbArena, we propose \textbf{SymbolicLM}, a framework that explicitly trains general-purpose LLMs as symbolic regressors. The framework consists of three stages. First, instruction tuning establishes the mapping from numerical observations to symbolic expressions. Second, reinforcement fine-tuning further aligns the model with symbolic regression objectives through physics-symbolic rewards, encouraging the generation of syntactically valid, physically consistent, and structurally accurate expressions. Finally, during inference, \textbf{SymbolicSGA} further improves equation recovery by iteratively refining candidate expressions based on numerical evaluation and symbolic feedback. The overall framework is illustrated in Figure~\ref{fig:overview}.

\subsection{Instruction Tuning}
Building on PhysSymbArena, we first perform supervised instruction tuning to establish the mapping from numerical observations to symbolic expressions. Each training sample consists of a set of numerical observations $\mathcal{D}_i$, associated physical descriptions $C_i$, and the corresponding symbolic expression $f_i^*$:
$
\mathcal{D}_i
=
\left\{
(\mathbf{x}_{i,k},y_{i,k})
\right\}_{k=1}^{N_i},
y_{i,k}=f_i^*(\mathbf{x}_{i,k}).
$
For each sample, the input prompt is constructed as:
$
P_i=[I;V_i],
$
where $I$ specifies the symbolic regression instruction and associated physical descriptions, while $V_i$ contains the numerical observations in a structured format:$
\texttt{x\_0 = \ldots, x\_1 = \ldots, f(x)=\ldots}.$
Following previous symbolic regression approaches~\citep{kamienny2022end}, we represent the target expression $f_i^*$ using prefix notation to facilitate sequence generation. For example,$
    f(x)=2\sin(x)+3
$
is represented as
$
    [\texttt{add mul 2 sin x 3}]
$. Let $z_i^{\ast}=(z_{i,1}^{\ast},\ldots,z_{i,T_i}^{\ast})$ denote the target token sequence. We optimize the model using the standard token-level cross-entropy loss:
\begin{equation}
    \mathcal{L}_{\mathrm{SFT}}
    =
    -
    \frac{1}{N_{\mathrm{train}}}
    \sum_{i=1}^{N_{\mathrm{train}}}
    \sum_{t=1}^{T_i}
    \log
    \pi_{\theta}
    \left(
        z_{i,t}^{\ast}
        \mid
        P_i,z_{i,<t}^{\ast}
    \right).
\end{equation}

\begin{figure*}[t]
    \centering
    \includegraphics[width=0.95\textwidth]{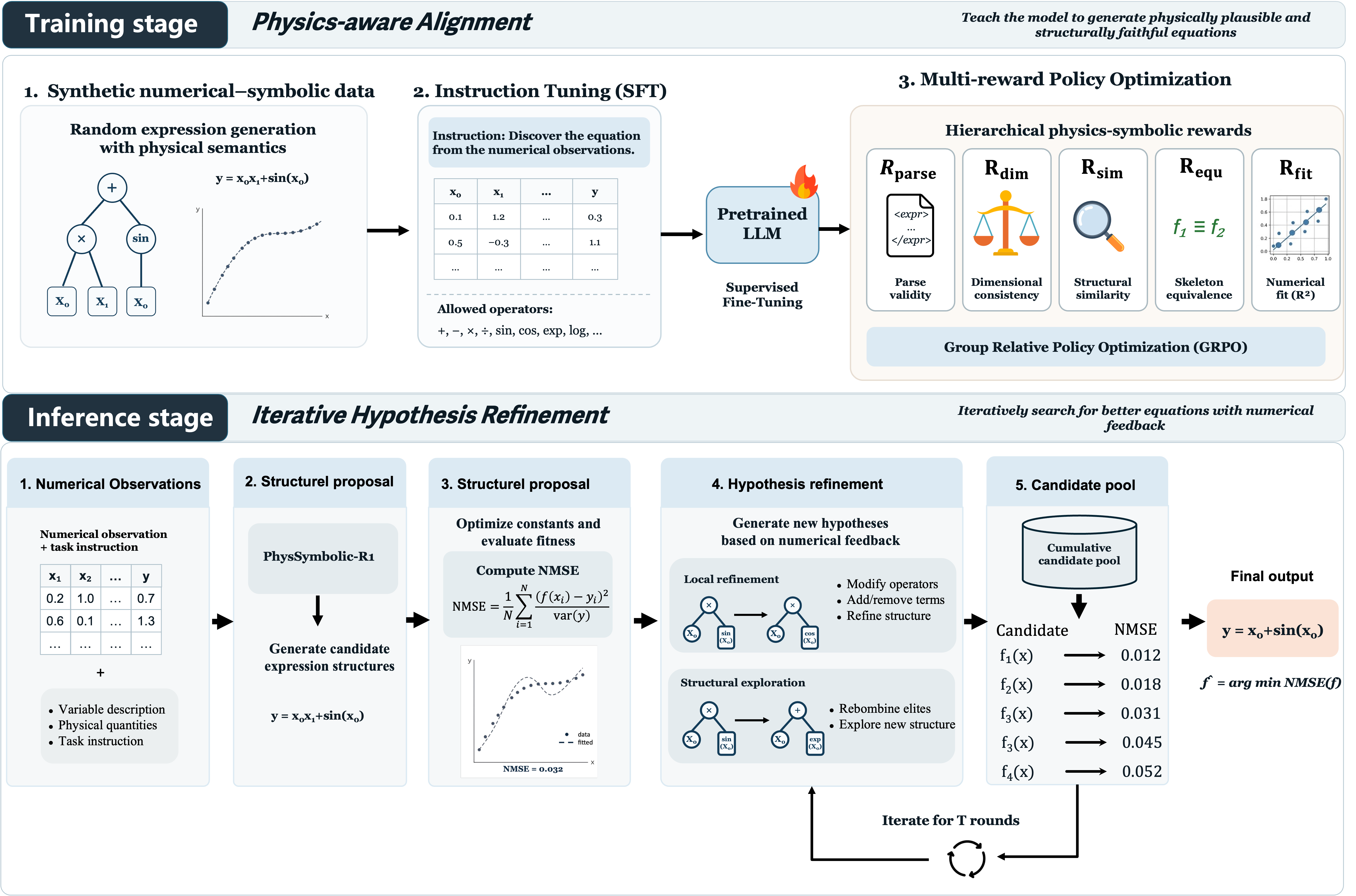}
    \caption{Overview of our proposed method.}
    \label{fig:overview}
\end{figure*}

\subsection{Physics-Symbolic Rewards}
\label{sec:rewards}
While instruction tuning establishes the mapping from numerical observations to symbolic expressions, token-level supervision alone cannot fully optimize symbolic regression objectives. We therefore further apply reinforcement fine-tuning with multiple rewards covering syntactic validity, physical consistency, symbolic structure, and numerical accuracy. Specifically, we introduce five complementary rewards: parse validity, dimensional consistency, structural similarity, symbolic equivalence, and numerical fitting. These rewards guide the model to generate valid, physically consistent, and accurate symbolic expressions.

\paragraph{Parse-Validity Reward.}
We define a parse-validity reward $R_{\mathrm{parse}}$ to evaluate whether generated expressions satisfy the required syntax and can be successfully parsed. Specifically, we introduce a validity checking function $\operatorname{isValid}$ and use SymPy to parse each generated expression. An expression that fails parsing is considered invalid and receives zero reward, while valid expressions receive a reward of one:
\[
R_{\mathrm{parse}}(z)
=
\begin{cases}
1, & \operatorname{isValid}(z)=1,\\
0, & \text{otherwise}.
\end{cases}
\]

\paragraph{Dimensional-Consistency Reward.}
To encourage the model to generate physically consistent expressions, we introduce a dimensional-consistency reward $R_{\mathrm{dim}}$. This reward leverages the physical information provided in PhysSymbArena and evaluates the dimensional consistency of generated expressions according to the validation rules introduced in Section 3. Expressions that violate dimensional constraints are considered invalid and receive a reward of zero, whereas expressions satisfying the constraints receive a reward of one:


\[
    R_{\mathrm{dim}}(\hat{f})
    =
    \begin{cases}
        1, & \operatorname{isDimValid}(\hat{f},y)=1,\\
        0, & \text{otherwise},
    \end{cases}
\]

\paragraph{Symbolic-Structure Similarity Reward.}
To provide dense structural supervision, we compare the  expression skeletons of the predicted and ground-truth expressions. 
Let $\mathcal{S}(\cdot)$ replace numerical constants except exponents with a shared placeholder $c$:
$
    s_{\hat{f}}=\mathcal{S}(\hat{f}),
    s^{\ast}=\mathcal{S}(f^{\ast}).
$ Each skeleton is characterized by
$
    \Phi(s)
    =
    \left(
        \mathcal{O}(s),
        \mathcal{F}(s),
        \mathcal{V}(s),
        N_c(s),
        P(s),
        C(s)
    \right),
$
where the components describe operators, nonlinear functions, variables, constants, structural sequence, and expression complexity. The structural similarity reward is
\[
    R_{\mathrm{sim}}(\hat{f},f^{\ast})
    =
    \sum_{m=1}^{6}
    w_m
    \operatorname{Sim}_m
    \left(
        \Phi_m(s_{\hat{f}}),
        \Phi_m(s^{\ast})
    \right),
    \qquad
    \sum_{m=1}^{6}w_m=1.
\]
Set-valued features are compared using Jaccard similarity, while count- and sequence-based features use normalized similarity measures.Since each component similarity is normalized to $[0,1]$. Therefore, even when a candidate expression recovers only a subset of the correct variables, operators, functions, or local structures, it receives a continuous reward proportional to its structural agreement with the ground truth. This feedback encourages the model to progressively approach the target analytical structure.

\paragraph{Numerical-fit reward.}
We use the numerical reward $R_{\mathrm{num}}$ to quantify how well an equation fits the input-output data. This reward is only utilized for equations passed the check of the valid decision function. We use the $R^2$ score to achieve the numerical reward. However, since the raw $R^2$ score is unbounded from below, its direct use can lead to training instability. We therefore apply a truncation strategy, defining the final reward as follows:
\[
    R_{\mathrm{num}}(\hat{f})
    =
    \begin{cases}
        \max\left(0,R^2(\hat{f},f^{\ast})\right),
        & \operatorname{isValid}(\hat{f})=1,\\
        0, & \text{otherwise}.
    \end{cases}
\]
This ensures that only syntactically correct equations are eligible for a numerical reward, while
invalid ones are penalized with a score of zero.

\paragraph{Symbolic-Equivalence Reward.}
We define an equivalence reward $R_{\mathrm{equ}}$ to encourage the model to recover expressions with symbolic structures equivalent to the ground-truth expressions. Since exact skeleton recovery is crucial yet challenging in symbolic regression, this reward provides positive feedback when the predicted expression and the ground-truth expression share the same skeleton structure. Specifically, it is defined as:
\[
R_{\mathrm{equ}}(\hat f_i,f_i)
=
\begin{cases}
1,&s_{\hat f_i}=s_{f_i},\\
0,&otherwise.
\end{cases}
\]
where $\hat{e}_i$ and $e_i$ denote the skeleton structures extracted from the predicted expression $\hat{f}_i$ and the ground-truth expression $f_i$, respectively. Specifically, numerical coefficients are replaced with shared placeholders to eliminate differences caused by constant values and focus on the symbolic structure.

\paragraph{Reinforcement Fine-tuning.}
The overall reward adopts a gated design. Parse validity and dimensional consistency serve as fundamental gates: if the generated expression fails either syntactic parsing or dimensional validation, the overall reward is set to zero. Otherwise, additional rewards for symbolic structure, symbolic equivalence, and numerical fitting are considered. Formally, the reward is defined as:
\[
R =
R_{\mathrm{parse}}R_{\mathrm{dim}}
(
w_{\mathrm{sim}}R_{\mathrm{sim}}
+
w_{\mathrm{equ}}R_{\mathrm{equ}}
+
w_{\mathrm{num}}R_{\mathrm{num}}
).
\]
we employ the Group Relative Policy Optimization (GRPO) algorithm~\citep{shao2024deepseekmath}. Specifically, for each input prompt, we sample a group of $G=8$ candidate expressions from the current policy. Each generated expression is evaluated using the proposed reward function, and the obtained rewards are normalized within the sampled group to compute relative advantages. These advantages are then used to update the policy, encouraging the model to generate symbolic expressions that satisfy syntactic validity, physical consistency, structural correctness, and numerical accuracy.



\subsection{SymbolicSGA}
\label{sec:inference}
During inference, we introduce \textbf{SymbolicSGA}, a lightweight external scaffold to maximize the model's symbolic regression capabilities. Our approach is inspired by SGA~\citep{ma2024llm}, a ReAct-based framework for scientific discovery that incorporates two distinct branches: exploitation and exploration. These two branches regulate search behaviors through generation temperatures, using low temperatures for exploitation to achieve local refinement and high temperatures for exploration to foster diversity. SGA then selects the top-5 candidates based on verification results  to feed back into the model for the next round of refinement.

We adapt SGA to symbolic regression by modifying both the verification metric and the branching mechanism. For each equation, we evaluate numerical fitness on the data as the verification metric and retain the top-5 candidates for iterative refinement. Rather than adjusting sampling temperatures, we employ task-specific prompts to guide the search. Specifically, the exploitation branch prompts the model to modify variable names or substitute unary and binary terms from the current best equation. Conversely, in the exploration branch, the model is prompted to generate new candidates distinct from the current worst equation while encouraging the reuse of sub-expressions from high-fitness candidates. By leveraging symbolic-aware structural priors, this prompt-guided refinement effectively balances local optimization with global functional exploration, enabling the model to converge to exact analytical equations within few iterations.

\section{Experiments}
\label{sec:experiments}

\subsection{Experimental Setup}

\paragraph{Datasets.}
We evaluate SymbolicLM on three symbolic regression benchmarks: PhysSymbArena, SRBench-Feynman~\citep{la2021contemporary}, and LLM-SRBench-Transform~\citep{shojaee2025llmsrbench}. 
PhysSymbArena is our proposed large-scale numerical--symbolic benchmark for training and evaluating LLM-based symbolic regression. 
SRBench-Feynman contains 119 physics equations from the Feynman equation dataset and is used to evaluate equation discovery on established physical problems. 
LLM-SRBench-Transform provides transformed symbolic regression tasks for evaluating the generalization ability of LLM-based methods beyond direct equation matching.


\paragraph{Baselines.}
We compare SymbolicLM with representative methods spanning conventional symbolic regression, neural symbolic regression, and LLM-based approaches. Conventional symbolic regression methods include AFP, AFP-FE, GP-GOMEA, GPLearn, and PySR, which perform explicit search over symbolic expression spaces. Neural symbolic regression methods, including DSR, E2E, SNIP, and TPSR, are included to compare with models specifically trained for numerical-to-symbolic mapping. We further include recent LLM-based symbolic regression methods, such as LLM-SR and SGA. In addition, general-purpose LLMs are evaluated to measure the symbolic regression capability of LLMs without task-specific adaptation. More details are provided in the Appendix.

\paragraph{Metrics.}
Following the evaluation protocols proposed in prior studies~\citep{kamienny2022end,biggio2021neural,la2021contemporary}, we adopt two widely used metrics to assess numerical accuracy: the coefficient of determination ($R^2$) and tolerance-based accuracy ($\text{Acc}_\tau$), which reports whether the worst-case relative error is within tolerance $\tau$, given by:
   \begin{equation}
 R^{2}=1-\frac{\sum_{i=1}^{N_{\text {test }}}\left(f(x_i)-\hat{f}(x_i)\right)^{2}}{\sum_{i=1}^{N_{\text {test }}}\left(f(x_i)-\overline{f(x_i)}\right)^{2}},\ \ \ \ 
 \mathrm{Acc}_{\tau} = \mathds{1} \left( \max_{1 \leq i \leq N_{\text{test}}} \left| \frac{\hat{f}(x_i) - f(x_i)}{f(x_i)} \right| \leq \tau \right),
\end{equation}
   where $\hat{f}(x_i)$ and $f(x_i)$ are the dependent outputs generated by feeding the same data into the predicted and ground-truth equations.

\begin{table}[t]
\centering
\caption{Results on PhysSymbArena.}
\label{tab:symbarena}
\small
\setlength{\tabcolsep}{6pt}
\renewcommand{\arraystretch}{0.82}
\begin{tabular}{lccc}
\toprule
Method
& $R^2 \uparrow$
& $\mathrm{NMSE} \downarrow$
& $\mathrm{Acc}_{\tau} \uparrow$ \\
\midrule

\multicolumn{4}{c}{\textbf{Specialized SR Methods}} \\
\midrule

DSR & 0.3239 & 0.6761 & 0.0237 \\
AFP-FE & 0.4078 & 0.5922 & 0.0830 \\
AFP & 0.3931 & 0.6069 & 0.0751 \\
E2E & 0.6297 & 0.3703 & 0.2727 \\
GP-GOMEA & 0.5468 & 0.4532 & 0.2925 \\
GPLearn & 0.2579 & 0.7421 & 0.0632 \\
SNIP & 0.8707 & 0.1293 & 0.3676 \\
TPSR & 0.7871 & 0.2129 & 0.3241 \\
PySR & 0.8652 & 0.1348 & 0.5059 \\

\midrule
\multicolumn{4}{c}{\textbf{General-purpose LLMs}} \\
\midrule

Qwen3-8B & 0.0179 & 0.9821 & 0.0119 \\
DeepSeek-V4-flash & 0.4218 & 0.5782 & 0.2946 \\
Gemini-3.8-flash & 0.4538 & 0.5462 & 0.3294 \\
GPT-5.5 & 0.6150 & 0.3889 & 0.3359 \\

\midrule
\multicolumn{4}{c}{\textbf{LLM-based SR Methods}} \\
\midrule

SGA (Qwen3-8B) & 0.4710 & 0.5290 & 0.1265 \\
LLM-SR (Qwen3-8B) & 0.6771 & 0.3229 & 0.2727 \\

\midrule
\multicolumn{4}{c}{\textbf{Ours}} \\
\midrule

SymbolicLM(SFT) & 0.7581 & 0.2496 & 0.4375 \\
SymbolicLM(SFT+RL) & 0.7626 & 0.2451 & 0.4648 \\
SymbolicLM + SymbolicSGA & \textbf{0.9056} & \textbf{0.1061} & \textbf{0.5508} \\

\bottomrule
\end{tabular}
\end{table}

\subsection{Results on PhysSymbArena}
Table~\ref{tab:symbarena} investigates whether symbolic regression can be acquired by general-purpose LLMs through dedicated numerical-symbolic training. Without task-specific adaptation, the pretrained Qwen3-8B model shows limited ability to recover analytical expressions, achieving only $R^2=0.0179$ and $\mathrm{Acc}_{\tau}=0.0119$. After supervised fine-tuning on PhysSymbArena, SymbolicLM(SFT) substantially improves equation recovery, reaching $R^2=0.7581$ and $\mathrm{Acc}_{\tau}=0.4375$, outperforming general-purpose LLM baselines and existing LLM-based symbolic regression approaches. These results indicate that numerical-symbolic supervision can effectively transform a general-purpose LLM into a model capable of performing symbolic regression.


We further investigate the effect of reinforcement learning alignment. Compared with SymbolicLM(SFT), incorporating GRPO with physic-symbolic rewards further improves structural recovery performance. Specifically, SymbolicLM(SFT+RL) increases exact accuracy from $0.4375$ to $0.4648$, while maintaining comparable numerical fitting performance. This improvement indicates that reward alignment provides additional supervision beyond token-level imitation, encouraging SymbolicLM to generate expressions that better satisfy symbolic structures and physical constraints. Finally, we evaluate the effect of inference-time refinement. Starting from SymbolicLM(SFT+RL), SymbolicSGA iteratively refines candidate equations using numerical feedback. After incorporating SymbolicSGA, SymbolicLM achieves$\mathrm{Acc}_{\tau}=0.5508$. Compared with SymbolicLM(SFT), the complete framework improves exact symbolic recovery accuracy by 11.33 \%. This demonstrates that quantitative feedback can effectively correct structurally inaccurate hypotheses and further improve equation discovery.

\subsection{Results on External Symbolic Regression Benchmarks}

Table~\ref{tab:sr_results} summarizes the performance of SymbolicLM on two external symbolic regression benchmarks, SRBench-Feynman and LLM-SRBench-Transform. These benchmarks evaluate whether the symbolic regression capability learned from PhysSymbArena can generalize beyond the training distribution to independently constructed equation discovery tasks.

On SRBench-Feynman, which contains physical equations from the Feynman dataset, SymbolicLM achieves the highest symbolic recovery accuracy among all compared methods, indicating its ability to recover underlying analytical structures rather than merely optimize numerical fitting. Meanwhile, it maintains competitive numerical accuracy compared with specialized symbolic regression approaches.

On LLM-SRBench-Transform, which evaluates symbolic regression under transformed equation representations, SymbolicLM also achieves strong performance. The results demonstrate that the model can infer underlying functional relationships from numerical observations beyond direct equation matching. Together, these results show that SymbolicLM effectively transfers the symbolic regression capability acquired from PhysSymbArena to diverse equation discovery scenarios.

\begin{table*}[t]
\centering
\caption{Results on SRBench-Feynman and LLM-SRBench-Transform.}
\label{tab:sr_results}
\small
\setlength{\tabcolsep}{5pt}
\renewcommand{\arraystretch}{0.92}

\begin{tabular*}{\textwidth}{@{\extracolsep{\fill}}lcccccc@{}}
\toprule
\multirow{2}{*}{Method}
& \multicolumn{3}{c}{SRBench-Feynman}
& \multicolumn{3}{c}{LLM-SRBench-Transform} \\
\cmidrule(lr){2-4}
\cmidrule(lr){5-7}
& $R^2 \uparrow$
& $\mathrm{NMSE}\downarrow$
& $\mathrm{Acc}_{\tau}\uparrow$
& $R^2 \uparrow$
& $\mathrm{NMSE}\downarrow$
& $\mathrm{Acc}_{\tau}\uparrow$\\
\midrule

AFP-FE
&0.6149&0.3850&0.1344
&0.4763&0.5237&0.0270\\

AFP
&0.6006&0.3993&0.0924
&0.4661&0.5339&0.0090\\

DSR
&0.6539&0.3460&0.0672
&0.5931&0.4069&0.0000\\

E2E
&0.9111&0.0888&0.4117
&0.7248&0.2752&0.1622\\

GP-GOMEA
&0.6018&0.3981&0.4033
&0.6308&0.3692&0.3333\\

GPLearn
&0.5821&0.4178&0.2184
&0.3279&0.6720&0.0270\\

SNIP
&\textbf{0.9683}&\textbf{0.0316}&0.3308
&0.8925&0.1075&0.1622 \\
TPSR
&0.9060&0.0939&0.5042
&0.8941&0.1059&0.2432\\

PySR
&0.9548&0.0451&0.6386
&0.9390&0.0610&0.3423\\

LLM-SR (Qwen-8B)
&-&-&-
&0.7919&0.2081&0.3333\\

SGA(Qwen-8B)
&-&-&-
&0.3200&0.6800&0.1441\\

\midrule

SymbolicLM + SymbolicSGA
&0.9569&0.0430&\textbf{0.6470}
&\textbf{0.9481}&\textbf{0.0519}&\textbf{0.4144}\\

\bottomrule
\end{tabular*}
\end{table*}

\section{Conclusion}
\label{sec:conclusion}
In this work, we introduce \textbf{PhysSymbArena} and \textbf{SymbolicLM} to equip LLMs with symbolic regression capabilities through numerical--symbolic and physical supervision. Experiments across multiple benchmarks demonstrate that SymbolicLM can effectively recover analytical structures from numerical observations. These results indicate that carefully designed numerical, symbolic, and physical supervision can effectively improve the symbolic regression capabilities of LLMs and provide a promising direction toward using language models for scientific equation discovery.


\subsection*{AI use statement}
In this work, we used generative AI tools for assistance with writing refinement, language editing, and improving the clarity and organization of the manuscript. We have not used generative AI tools for generating research results, designing experiments, conducting scientific analysis, or making methodological decisions. All AI-assisted outputs were reviewed and verified by the authors. The authors manually checked the correctness of the code, experimental configurations, mathematical formulations, and all reported results. The authors also reviewed the related literature to ensure the originality and proper attribution of the ideas discussed in this work. We take responsibility for the final content of this work, including text, claims or artifacts produced with the aid of generative AI.
\subsection*{Ethics statement}
The authors acknowledge their responsibility to adhere to the ICLR Code of Ethics.
\subsection*{Reproducibility statement}
To ensure the reproducibility of our results, we provide comprehensive details of our methodology, experiments, and implementation. A full description of our model architectures, algorithms, and experimental setup is provided in Appendix. We believe this provides sufficient information for the research community to reproduce and build upon our findings.

\bibliography{iclr2027_conference}
\bibliographystyle{iclr2027_conference}

\end{document}


\appendix
\section{Appendix}

\subsection{Prompt Detail}
\begin{tcolorbox}[
colback=gray!5,
colframe=gray!30,
title=System Prompt
]

 You are a leading physicist and AI algorithms expert, skilled at Physics-Informed Symbolic Regression. Your task is to derive realistic and high-accuracy mathematical formulas from the given [Target variable], [Available feature variables] with their physical dimensions, and
  [Observed data points], and output the formula in the required format.\\
  === Core reasoning rules ===\\
  1. Dimensional consistency:\\
     - Both sides of addition/subtraction (+, -) must have exactly the same dimensions.\\
     - Arguments of transcendental functions (sin, cos, exp, log, etc.) must be dimensionless.\\
  2. Variable combination first:\\ Prefer combining available feature variables $(x_i)$ through multiplication/division to match the dimensions
  of the target variable (y).
\end{tcolorbox}

\begin{tcolorbox}[
colback=gray!5,
colframe=gray!30,
title=User Prompt
]
 === Input information ===\\
  {Target variable $y$}\\
  - Name: {target name} {target symbol}\\
  - Physical dimension: {target dimension}\\
{Available feature variables $X$}\\
  - $x_0$: {name}, Physical dimension: {dimension}\\
  - $x_1$: {name}, Physical dimension: {dimension}\\
  ...\\
  {Observed data points $->$ $y$}\\
  $x_0$={value}, $x_1$={value}, ..., $y$={value}\\
  ..200 data point pairs...\\

  === Output requirement ===\\
  Output only one final mathematical expression in standard infix notation, wrapped exactly as $<expr>$...$</expr>$. Do not use prefix notation and do not
  include any explanation.\\

  Use variables $x_0$, $x_1$, $...$
  Supported operators: +, -, *, /, **, sin, cos, tan, asin, acos, atan, exp, log, sqrt, abs.
  Use abs for absolute value. Do not include any explanation.\\
\end{tcolorbox}

\begin{tcolorbox}[breakable,
    colback=gray!5,
    colframe=gray!30,
    title=Local Exploitation Prompt
]

\raggedright

=== Local exploitation: iteration [iteration] ===\\[4pt]

\textit{[variable descriptions]}\\[4pt]

The formula below is the mandatory parent selected because it has the lowest NMSE.\\
Do not treat it as a loose reference and do not regenerate a formula from scratch.\\[4pt]

Best parent formula:\\
\texttt{[parent formula]}\\[4pt]

Parent NMSE (lower is better):\\
\texttt{[parent NMSE]}\\[4pt]

Generate one local structural variant of the parent.\\[4pt]

Requirements:\\[2pt]
1. Preserve the parent's useful main structure and variable combination.\\
2. Make exactly one small structural modification.\\
3. Prefer to \texttt{[mutation operation]}.\\
4. Do not change only numerical constants; constants will be optimized by BFGS.\\
5. Do not introduce unrelated variables or rebuild the entire expression.\\
6. Keep the expression physically meaningful and dimensionally consistent when dimensions are available.\\[4pt]

Observed data points $\rightarrow y$:\\
\textit{[observed data points]}\\[4pt]

=== Output requirement ===\\[2pt]

Output only the final prefix expression that fits the data.\\
Use variables \texttt{[available variables]}.\\
Do not include any explanation.

\end{tcolorbox}

\begin{tcolorbox}[breakable,
colback=gray!5,
colframe=gray!30,
title=Structural Exploration Prompt
]

=== Local exploitation: iteration [iteration] ===\\[4pt]
\textit{[variable descriptions]}\\[4pt]
The candidate formulas below are ordered from best to worst by NMSE.\\
Candidate [0] is the current best formula.\\[2pt]
Top candidate formulas:\\
\textless formula 0\textgreater \quad (NMSE=\textless nmse 0.01\textgreater)\\
\textless formula 1\textgreater \quad (NMSE=\textless formula 1\textgreater \quad (NMSE=\\textless nmse 0.03\textgreater)\\nmse 0.03\>)\\
\vdots\\
\textless formula k\textgreater \quad (NMSE=\textless formula 1\textgreater \quad (NMSE=\\textless nmse 0.03\textgreater)\\nmse 0.091\textgreater)\\[2pt]
Generate a structurally exploratory formula by recombining useful subexpressions from these candidates.\\[2pt]
Requirements:\\
1. Retain at least one useful variable combination or subtree from the top candidates.\\
2. Explore a meaningfully different structure rather than merely perturbing numerical constants.\\
3. Prefer substructures shared by multiple good candidates and avoid structures found only in poor candidates.\\
4. Keep the expression physically meaningful and dimensionally consistent when dimensions are available.\\
5. Output one formula only.\\[2pt]

\textit{[observed data points]}\\[4pt]
=== Output requirement ===\\
Output only the final prefix expression that fits the data.\\
Do not include any explanation.

\end{tcolorbox}
\subsection{Baseline}
We compare SymbolicLM with representative methods spanning conventional symbolic regression, neural symbolic regression, and LLM-based approaches. Conventional symbolic regression methods include AFP~\citep{schmidt2010age}, AFP-FE~\citep{schmidt2009distilling}, GP-GOMEA~\citep{virgolin2021improving}, GPLearn~\citep{schmidt2010age}, and PySR~\citep{cranmer2023interpretable}, which perform explicit search over symbolic expression spaces. Neural symbolic regression methods, including DSR~\citep{landajuela2022unified}, E2E~\citep{kamienny2022end}, SNIP~\citep{meidani2024snip}, and TPSR~\citep{schmidt2010age}, are included to compare with models specifically trained for numerical-to-symbolic mapping. We further include recent LLM-based symbolic regression methods, such as LLM-SR~\citep{shojaee2025llm} and SGA~\citep{shojaee2025llmsrbench}. For LLM-SR, the maximum number
of sampled equations is capped at 32. For SGA, the process is configured for 5 iterations, with
each iteration involving the exploitation of two equations and the exploration of four. The LLM backbone is Qwen3-8B. In addition, general-purpose LLMs are evaluated to measure the symbolic regression capability of LLMs without task-specific adaptation.

\subsection{Comparison between SymbolicSGA and SGA}

We adapt SGA to symbolic regression by modifying both the verification metric and the branching mechanism. To evaluate the effectiveness of our proposed SymbolicSGA refinement framework, we compare it with the original SGA-based refinement strategy under the same model and dataset settings on the SRBench-Feynman dataset. The results are shown in Table~\ref{tab:sga_comparison}.

\begin{table}[h]
\centering
\caption{Comparison between SymbolicSGA and SGA refinement strategies.}
\label{tab:sga_comparison}
\small
\begin{tabular}{lccc}
\toprule
Method & $R^2 \uparrow$ & NMSE $\downarrow$ & ACC $\uparrow$ \\
\midrule
SGA & 0.9539 & 0.0461 & 0.6218 \\
SymbolicSGA & \textbf{0.9569} & \textbf{0.0431} & \textbf{0.6471} \\
\bottomrule
\end{tabular}
\end{table}

Compared with the original SGA refinement strategy, SymbolicSGA achieves consistent improvements across all evaluation metrics. Specifically, SymbolicSGA increases $R^2$ from 0.9539 to 0.9569, reduces NMSE from 0.0461 to 0.0431, and improves exact symbolic recovery accuracy from 0.6218 to 0.6471. These results demonstrate that the proposed SymbolicSGA is better suited for symbolic regression tasks. By incorporating symbolic-aware refinement strategies and equation-specific feedback, SymbolicSGA can more effectively revise candidate expressions and recover accurate analytical structures from numerical observations.

\subsection{Effect of Physical Information in PhysSymbArena}
\begin{table}[h]
\centering
\caption{Effect of Physical Information.}
\label{tab:physical}
\small
\begin{tabular}{lccc}
\toprule
Method & $R^2 \uparrow$ & NMSE $\downarrow$ & ACC $\uparrow$ \\
\midrule
Without Physical Information & 0.7678 & 0.2400 & 0.4297 \\
With Physical Information & 0.7582 & 0.2497 & \textbf{0.4375} \\
\bottomrule
\end{tabular}
\end{table}

To investigate the contribution of physical information in PhysSymbArena, we conduct an ablation study by training SymbolicLM with and without physical descriptions, including variable semantics and dimensional information. The results are summarized in Table~\ref{tab:physical}.

The results show that physical information does not directly improve numerical fitting performance, with slightly lower $R^2$ and higher NMSE. However, it improves symbolic recovery accuracy, increasing ACC from 0.4297 to 0.4375. This indicates that physical information mainly provides semantic and structural guidance, helping SymbolicLM recover more physically meaningful symbolic expressions.

\subsection{Physical Grounding and Dimensional Validation}
To endow abstract expressions with real-world physical semantics, we curate a dictionary of 127 physical quantities from the SymPy library, covering multiple domains such as mechanics, electromagnetism, and thermodynamics. For each expression, physical quantities are randomly sampled from the same domain and assigned to the input variables $\mathbf{x}$ and the target output $y$. We propagate dimensional information recursively along the expression tree according to SI rules. For addition and subtraction, the operands must have identical dimensions:
$ \mathbf{d}(a\pm b)=\mathbf{d}(a), \text{where } \mathbf{d}(a)=\mathbf{d}(b).$
For multiplication and division, the dimensions are propagated as:
$\mathbf{d}(ab)=\mathbf{d}(a)+\mathbf{d}(b),
\mathbf{d}\left(\frac{a}{b}\right)=\mathbf{d}(a)-\mathbf{d}(b).$
For exponentiation with a constant exponent $r$, the resulting dimension is:
$\mathbf{d}(a^r)=r\mathbf{d}(a).$
The arguments of transcendental functions, including $\sin$, $\exp$, and $\log$, must be dimensionless:
$\mathbf{d}(a)=\mathbf{0}$. Any assignment violating dimensional consistency is discarded and resampled, ensuring that the generated physical metadata $\mathcal{C}$ satisfies fundamental physical constraints.

\bibliography{iclr2027_conference}
\bibliographystyle{iclr2027_conference}